\documentclass[12pt]{article}
\usepackage{amsfonts}

\newcommand{\be}{\begin{equation}}
\newcommand{\ee}{\end{equation}}
\newcommand{\ba}{\begin{eqnarray}}
\newcommand{\ea}{\end{eqnarray}}

\begin{document}

\begin{titlepage}
\rightline{{CPHT RR 042.0703}}
\rightline{{LPT-ORSAY 03-57}}
\rightline{hep-th/0309057}

\vskip 2cm
\centerline{{\large\bf On cosmologically induced hierarchies in string theory }}
\vskip 1cm
\centerline{E. Dudas${}^{\dagger,\star}$,
J. Mourad${}^{\star,*}$ and C. Timirgaziu${}^{\star}$}
\vskip 0.5cm
\centerline{\it ${}^\dagger$ Centre de Physique
  Th{\'e}orique\footnote{Unit{\'e} mixte du CNRS et de l'EP, UMR 7644.} ,
Ecole Polytechnique, F-91128 Palaiseau}
\vskip 0.3cm
\centerline{\it ${}^\star$ Laboratoire de Physique 
Th{\'e}orique
\footnote{Unit{\'e} Mixte de Recherche du CNRS (UMR 8627).}}
\centerline{\it Universit{\'e}  Paris XI, B{\^a}t. 210, F-91405 Orsay Cedex}
\vskip 0.3cm 
\centerline{\it ${}^*$ F{\'e}d{\'e}ration de Recherche APC,
Universit{\'e}  Paris VII,\\
2, place Jussieu, 75251 Paris}

\vskip  1.0cm
\begin{abstract}

We propose, within a perturbative string theory example, a cosmological
way to generate a large hierarchy between the observed Planck mass and the
fundamental string scale. Time evolution results in three large space
dimensions, one additional dimension transverse to our world and five 
small internal dimensions with a very slow time evolution. The evolution
of the string coupling and internal space generate a large Planck
mass. However due to an exact compensation between the time
evolution of the internal space and that of the string coupling, the
gauge and Yukawa couplings on our Universe are time independent.  

\end{abstract}
\end{titlepage}

%%%%%%%%%%%%%%%%%%%%%%%%%%%%%%%%%%%%%%%%%%%%%%%%%%%%%%%%%%%%%%%%%%%%%%%%%%%%%%%%%%%%%%%%%%%%%%%%%%
\section{Introduction}

Long time ago, Dirac initiated an ambitious  program to
relate very large numbers to the age of the universe
\cite{Dirac:1937ti}. This had as an immediate consequence, the 
variation in time of the
fundamental constants on which stringent experimental 
limits exist \cite{Uzan:2002vq}, ruling out many of the proposed models including
the simple model Dirac had himself proposed.
In addition to the well-known hierarchy
between the Planck scale on one hand and the electroweak scale,
and the  cosmological constant on the other, string theory
faces another similar problem which amounts 
to the explanation of the smallness of the extra dimensions.

Here, we report on a cosmological solution, 
obtained within an effective field theory approximation, 
which allows to obtain, {\it {\`a} la} Dirac, a large hierarchy between the  
Planck scale and the string scale as well as three large space dimensions,
a small compact five dimensional space and one large dimension transverse 
to our world. The model has the additional desirable feature
of having a perturbative string coupling
while maintaining a Yang-Mills coupling of order one.
The model is to be thought of as
a toy model allowing to obtain initial conditions 
in the primordial universe with the required hierarchies. So, as inflation models,
it is not intended to be realistic
until the present era. We do not address the difficult
problem of explaining why the hierarchies are maintained 
after this era. Neither  do we explain 
why we started with the particular string model
with the given supersymmetry breaking we consider.

This Letter is  organised as follows. 
In Section 2 we present the
string model as well as the cosmological background, obtained by
compactifying to four dimensions the solution found in \cite{dmt}.
Section 3 is devoted to the evaluation of the various scales and
couplings as functions of time. Section 4 discusses briefly 
the stability of the model and also shows that scalar fields
describing the position of branes in our cosmological background
get scalar potentials becoming naturally flat by the time evolution
and are therefore potential candidates for quintessential models of
dark energy.

%%%%%%%%%%%%%%%%%%%%%%%%%%%%%%%%%%%%%%%%%%%%%%%%%%%%%%%%%%%%%%%%%%%%%%%%%%%%%%%%%%%%%%%%%%%%%%%%%%%%
\section{A cosmological solution in string theory}

We are considering here a class of orientifold
string models \cite{augusto1,orientifold}
containing D8 branes and non-dynamical orientifold O8 planes, whose
specific charges will be discussed in two explicit models at the end
of this section. Supersymmetry is mainly broken on the D8/O8
system \cite{bsb}, while in the closed (bulk) sector is
either exactly supersymmetric or has softly broken supersymmetry in
the large radius limit \cite{ads}.  
The corresponding bosonic effective action is  
\ba
S &\!\!=\!\!& {M_s^8 \over 2 } \int d^{10} x \sqrt {-g} \biggl[ e^{-2 \Phi} (
R+ 4 ({\partial \Phi})^2)- {1 \over 2 \times 10 !} \ F_{10}^2 \biggr] 
\nonumber \\
&\!\!\!-\!\!\!& \int_{y=0} d^9 x (T_0 \ \sqrt{-\gamma} \ e^{-\Phi}
\ + q_0 \ A_9) \!-\! \int_{y= \pi R} d^9 x \ \sqrt{-\gamma} \ 
[e^{-\Phi} (T_1+ {1 \over g^2} {\rm tr} F^2) + q_1 A_9 ] \ , \label{s1}
\ea
where $M_s$ is the string scale, $\Phi$ is the dilaton, $A_9$ the RR nine-form coupled to D8
branes and O8 planes, $\gamma$ is the induced metric and $(1/g^2) \sim
M_s^5$ is related to the 9d Yang-Mills dimensionful coupling. A certain
number of D8 branes of NS-NS and RR charges $(T_0,q_0)$ were placed at the origin
$y=0$ and the rest of them, containing gauge fields $F$, were
placed at $y=\pi R$ of a compact coordinate $y$ of radius $R$. 
An important fact for the following is that whereas the model has no
Ramond-Ramond tadpoles by consistency reasons, it has dilaton
tadpoles \cite{bsb}. In the generic case 
\be
T_0^2 \ge q_0^2 \quad , \quad T_1^2 \ge q_1^2 \ , \label{s2}
\ee
it was shown in \cite{dmt} that the field equations of (\ref{s1}) result
in a space-time metric and string coupling 
\ba
ds^2 &=&  \biggl[ G_0+ {3 |q_0| {\kappa}^2 \over 2 \lambda} e^{
\lambda t} sh (\lambda |y|) \biggr]^{-{1 \over 3}} \ \biggl[  
\delta_{\mu \nu} dx^{\mu} dx^{\nu} + e^{2 \lambda t} (-dt^2+dy^2) \biggr]
\ , \nonumber \\
e^{\Phi} &=&  \ \biggl[ G_0+ {3 |q_0| {\kappa}^2 \over 2 \lambda} e^{
\lambda t} sh (\lambda |y|) \biggr]^{- {5 \over 6}} \ , 
 \label{s3}   
\ea
whereas the Einstein frame metric $ds^2 = \exp (\Phi / 2) ds_E^2$ is
\be
ds_E^2 =  \biggl[ G_0+ {3 |q_0| {\kappa}^2 \over 2 \lambda} e^{
\lambda t} sh (\lambda |y|) \biggr]^{1 \over 12} \ \biggl[  
\delta_{\mu \nu} dx^{\mu} dx^{\nu} + e^{2 \lambda t} (-dt^2+dy^2)
\biggr] \ . \label{s03}
\ee
Here $G_0$ is an integration constant determining the string
coupling constant at the origin, $\kappa^2=1/M_s^{8}$
and $\lambda$ is determined below.
The solution (\ref{s03}) is nonsingular in the compact $y$
coordinate\footnote{For previous classical solutions of
nonsupersymmetric strings, see e.g. \cite{dm1}.}
and has big-bang type singularities at $t = \pm \infty$, separated by
an infinite proper time.
Notice that the string coupling is finite in the whole spacetime,
including at the big-bang singularities, rendering such solutions
attractive string frameworks for studying the pre big-bang
\cite{veneziano} and the epkyrotic \cite{epkyrotic} scenarios.   
The boundary conditions at the position of the branes
\be
T_0 = q_0 \quad , \quad {T_1 \over ch (\pi \lambda R)} = q_1 \  \label{s4} 
\ee
determine the parameter $\lambda$. 

By making the change of variables
\be
X_0 \ = {1 \over \lambda} \ e^{\lambda t} \ ch (\lambda y ) \quad , \quad 
X \ = {1 \over \lambda} \ e^{\lambda t} \ sh (\lambda y )  \ , \label{s5} 
\ee
we get the spacetime metric
\be
ds^2 =  
\biggl[ G_0+ {3 |q_0| {\kappa}^2 \over 2 } |X|   \biggr]^{-{1 \over 3}} \biggl[  
\delta_{\mu \nu} dx^{\mu} dx^{\nu} -dX_0^2+ dX^2 \biggr]
\ , \label{s6}   
\ee
when $y>0$. The $Z_2$ identification
$y\rightarrow -y$ is mapped in terms of the coordinates $(X_0,X)$
to the parity $\Pi_X$. This means that
the orientifold operation acts in the $(X_0,X)$ plane as 
$\Omega' \ = \ \Omega \ \Pi_X$.  
In addition, the identification of points on the circle $y = y+2 \pi
R$ results in $(X_0,X)$ coordinates in the orbifold identification 
\ba 
\left( 
\begin{array}{c} 
X_0 \\ 
X 
\end{array} 
\right)  
& \ \rightarrow \ & 
\left(
\begin{array}{cc} 
ch (2 \pi \lambda R) & sh (2 \pi \lambda R)  \\ 
sh (2 \pi \lambda R)  & ch (2 \pi \lambda R) 
\end{array}
\right)
\left(
\begin{array}{c}
X_0 \\
X
\end{array}
\right) \ , \label{s7}
\ea
which is nothing but a two-dimensional boost ${\cal K}_{2 \pi \lambda R}$ with a velocity 
$V = th(2 \pi \lambda R)$ in the $(X_0,X)$ space\footnote{For
the quantization of string models
on lorentzian orbifolds, see e.g. \cite{lorentzian}.}.
  
Notice that (\ref{s6}) coincides with the spacetime metric 
obtained in \cite{pw} in the supersymmetric
Type I' string with N D8 branes at the origin $X=0$ of a compact coordinate of
radius $R$ and $32-N$ D8 branes at $X=\pi R$~.

The metric (\ref{s6}) and the identifications $\Omega'$
and (\ref{s7}) allow a  simple physical 
interpretation of our configuration in the $(X_0,X)$ coordinates. 
Indeed, the fixed points of the two orientifold operations are
\be
\Omega'~: \  X = \ 0 \quad , \quad 
\Omega'~ {\cal K}_{2 \pi \lambda R} : \  X = th (\pi \lambda R ) \ X_0 \ . \label{s8}
\ee
Consequently, the branes (and orientifolds) located at the origin stay at the origin $X=0$
in the static background (\ref{s6}), whereas the
branes and the orientifold planes at $y=\pi R$ move at a
constant velocity $v_1=th \ (\pi\lambda R)$.
Moreover, the boundary conditions (\ref{s4}) encode the dynamics
of the two boundaries in the form
\ba
T_0 = q_0 \quad , \quad T_1 \sqrt {1-v_1^2} = q_1 \ , \label{s10}  
\ea
whose interpretation is quite obvious, since $T_1 \sqrt {1-v_1^2}$ is
the boosted tension of the branes/O-planes moving with the velocity
$v_1$ in the static background (\ref{s6}). The RR charge conservation
implies therefore the boosted version of the NS-NS tadpole conditions 
\be
T_0 \sqrt{1-v_0^2} + T_1 \sqrt{1-v_1^2} = 0 \ , \label{s11} 
\ee
where in our case $v_0=0$. 

These time-dependent solutions 
can be used to find four-dimensional cosmological solutions if 
we consider that all dimensions, except the
time and three noncompact space ones parallel to the branes, are
compact. For this purpose we
consider five space dimensions to span a five-torus and
toroidally compactify our solution in order to have D3-O3 systems
propagating into a five dimensional bulk space. 
We make the split $M = (\alpha, m)$, where $\alpha = 0 \cdots 4$ are
the noncompact spacetime coordinates plus the $X$ coordinate and $m = 5 \cdots 9$ are compact
coordinates parallel to the 8-branes/orientifold planes. 

Compactification from 10d to 5d in the string frame asks for the
relations
\be
g_{mn}^{(10)} = V_5^{2 \over 5} \delta_{mn} \quad , \quad 
g_{\alpha \beta}^{(10)} =  g_{\alpha \beta}^{(5)} \ , \label{s12}
\ee  
where $V_5 \equiv v_5 \exp (5 {\tilde \sigma})$ is the volume of the internal
five-torus with $v_5 \equiv r_c^5$ the constant volume parameter and 
${\tilde \sigma}$ is the breathing mode of the five-torus. In our case
\be
\langle V_5 \rangle \ = v_5 \ e^{5 <{\tilde \sigma}>} \ = \ v_5 \ 
\biggl[ G_0+ {3 |q_0| {\kappa}^2 \over 2 } |X| \biggr]^{-{5 \over 6}} \ , \label{s13}
\ee 
where the five-dimensional metric takes the form 
\be
ds_{5}^2 =  \biggl[ G_0+ {3 |q_0| {\kappa}^2 \over 2 } |X| 
\biggr]^{-{1 \over 3}} \ \biggl[  
\delta_{\mu \nu} dx^{\mu} dx^{\nu} -dX_0^2 + dX^2 \biggr] \ . \label{s14}  
\ee
It is useful for later purposes to display also the results in the 5d
Einstein frame. This can be achieved by the Weyl rescalings
\be
V_5  = e^{5 \Phi \over 4} V_{E,5} \quad , \quad 
g_{\alpha \beta}^{(5)} = e^{{4 \Phi \over 3}-{10 {\tilde \sigma} \over 3}} 
g_{E, \alpha \beta}^{(5)} \ . \label{s15}
\ee  
In the Einstein frame
\be
\langle V_{E,5} \rangle \ = \ v_5 \ 
\biggl[ G_0+ {3 |q_0| {\kappa}^2 \over 2 } |X| \biggr]^{5 \over 24} \ , \label{s16}
\ee 
and the five-dimensional metric takes the form 
\be
ds_{E,5}^2 =  \biggl[ G_0+ {3 |q_0| {\kappa}^2 \over 2 } |X| \biggr]^{2 \over 9} \ \biggl[  
\delta_{\mu \nu} dx^{\mu} dx^{\nu} -dX_0^2 + dX^2
\biggr] \ . \label{s117}  
\ee
This five-dimensional Einstein background is a classical solution of the
five-dimensional lagrangian
\ba
S_5 &=& {1 \over 2 {\kappa}_5^2} \int d^{5} x \sqrt {-g} \biggl[ 
R^{(5)} - {1 \over 2} ({\partial \Phi})^2 - {40 \over 3} ({\partial
\sigma})^2 - {1 \over 2 \times 5 !} \ e ^{- {5 \Phi \over 2} + {10
\sigma \over 3}} \ F_{5}^2 \biggr] \nonumber \\
&-& \int_{X=0} d^4 x \biggl[ \sqrt{-\gamma} \ 
 T_0 \ e^{ {5 \Phi \over 4} - {5 \sigma \over 3} } \ + q_0 \ A_4 + \cdots
\biggr]  \nonumber \\
&-& \int_{X=v_1 T} d^4 x \biggl[ \sqrt{-\gamma} \ 
 (T_1 \ e^{ {5 \Phi \over 4} - {5 \sigma \over 3} }+ 
{v_5 \over g^2} e^{{\Phi \over 4} + 5 \sigma } {\rm tr} F^2) \ + q_1 \ A_4 + \cdots
\biggr]
\ , \label{s17}
\ea
where $\sigma = {\tilde \sigma} - {\Phi /4}$ and $(1/ {\kappa}_5^2) =
v_5 M_s^8$ is obtained by a straightforward
dimensional reduction from the original 10d/9d one. 
In (\ref{s17}), 
$T_{0,1}$ $(q_{0,1})$ denote now the D3 branes tensions (RR charges) on
the two boundaries and $\cdots$ denote contributions of the internal 
components of the gauge fields $F$, as well as the fields describing 
the fluctuations of the branes.

If the time evolution drives the internal five-space to a size smaller than
the string scale, then we must perform T-duality in 
(\ref{s13}), (\ref{s14}) along the internal five-torus. The T-dual five
volume reads $V_5' = 1 / (V_5 M_s^{10})$ and the T-dual string coupling
$\exp (\Phi') = \exp (\Phi)/ (V_5M_s^5)$ is constant. The T-dual
solution describes smeared D3 branes along the five-torus. If the X
coordinate is much larger than the other five internal ones, there is a 
five dimensional T-dual lagrangian describing this solution, which reads 
\ba
S_5 &=& {1 \over 2 {\kappa}_5^2} \int d^{5} x \sqrt {-g} \biggl[ 
R^{(5)} - {1 \over 2} ({\partial \Phi'})^2 - {40 \over 3} ({\partial
\sigma'})^2 - {1 \over 2 \times 5 !} \ e ^{ {40
\sigma' \over 3}} \ F_{5}^2 \biggr] \nonumber \\
&-& \int_{X=0} d^4 x \biggl[ \sqrt{-\gamma} \ 
 T_0 \ e^{- {20 \sigma' \over 3} } \ + q_0 \ A_4 + \cdots
\biggr]  \nonumber \\
&-& \int_{X=v_1 T} d^4 x \biggl[ \sqrt{-\gamma} \ 
 (T_1 \ e^{- {20 \sigma' \over 3}}+ 
{v_5 \over g^2} e^{-\Phi'} {\rm tr} F^2) \ + q_1 \ A_4 + \cdots
\biggr] \ . \label{s18}
\ea
The T-dual fields $\Phi', \sigma'$ in the 5d Einstein frame  
are related to the ones before T-duality by 
$(\Phi', \sigma') = (-(\Phi /4)- 5 \sigma, (\sigma /4)- 3(\Phi /16))$.
   
%%%%%%%%%%%%%%%%%%%%%%%%%%%%%%%%%%%%%%%%%%%%%%%%%%%%%%%%%%%%%%%%%%%%%%%%%%%%%
\subsection{Explicit string examples}

Our first example is the model discussed in \cite{dmt}, which is an orientifold
\cite{augusto1,orientifold} of Type II
strings on ${\cal M}^9 \times S^1$ with broken supersymmetry \cite{ads}. 
The explicit construction starts from a freely-acting
orbifold $g$ in the closed sector of the type II string. After a radius redefinition, 
the orbifold $g$ becomes a periodic identification $y = y + 2 \pi R$
accompanied by the spacetime fermion number operation, imposing
different boundary conditions for bosons and fermions and breaking
therefore the supersymmetry. The orientifold operations $\Omega' = \Omega
\ \Pi_y \ (-1)^{f_L}$ and $\Omega' \ g$ create O-planes of two
different types. The fixed plane of $\Omega'$ sits at the origin $y=0$
and is a standard $O_{+}$ plane, whereas the fixed plane of $\Omega' \
g$ sits at $y = \pi R$ and is an antiorientifold plane, due to the
action of $(-1)^F$. More precisely, it is an  ${\overline O}_{-}$ plane
due to the simultaneous action of $(-1)^F$ and  $(-1)^{f_L}$
operations. The  $O_{+}-{\overline O}_{-}$ has the additional interesting and
important (for our purposes) property to eliminate the would be closed
string tachyon, which is odd under the world-sheet fermion number
operator  $(-1)^{f_L}$.  The model contains by consistency also 32 D8 branes and therefore no
open string tachyons exist.
The partition function for this non-tachyonic model and the standard
consistency checks were displayed in the Appendix of \cite{dmt}. 
The conditions (\ref{s2}) are satisfied in this explicit
example\footnote{At the one-loop level a cosmological constant
$\Lambda_1$ is generated in the bulk. In the large radius limit
$R M_s >>1$, relevant for our cosmological solution,
$\Lambda_1$ is however parametrically very small and can be
neglected.}, since
\be
T_0 = q_0 \ = \ (N-16) \ T_8 \quad , \quad T_1 \ = \ (48-N) \ T_8 \quad , \quad
q_1 \ = \ (16-N) \ T_8 \ . \quad \label{s04}
\ee

The second example starts with an orientifold of Type IIB superstring by
$\Omega' = \Omega \delta$, where $\delta y = y + \pi R$ is a
shift by half of the $S^1$ circle described by $y$. After T-duality on
$y$ the shift operation generates $O8_{+}$ planes at $y=0$ and
$O8_{-}$ planes at $y=\pi R$. The total RR charge being zero, the
orientifold does not ask for consistency any addition of D8 branes and
is a well known example of orientifolds without open strings
\cite{dp}. Let us now add, consistently with RR tadpole
cancellation, $N$ D8 branes at $y=0$ and an equal number $M=N$ of anti D8
branes at $y = \pi R$. In the large radius limit the would be open
string tachyons stretched between branes and antibranes are very massive and do not
play an important role in the dynamics of the system. Supersymmetry is
broken at the string scale at $y = \pi R$, whereas at the lowest order
it is exact in the bulk. There is an overall dilaton tadpole and the
bosonic effective action is again of the form (\ref{s1}), with the
obvious replacement $32-N \rightarrow M$ for the $y = \pi R$ localized
action. The various tensions and charges are in this case  
\be
T_0 = q_0 \ = \ (N-16) \ T_8 \quad , \quad T_1 \ = \ (N+16) \ T_8 \quad , \quad
q_1 \ = \ (16-N) \ T_8 \ . \quad \label{s19}
\ee
If the number of brane-antibrane pairs is $N < 16$, then the
conditions (\ref{s2}) are again fulfilled and the classical solution
(\ref{s3}) is valid.      
%%%%%%%%%%%%%%%%%%%%%%%%%%%%%%%%%%%%%%%%%%%%%%%%%%%%%%%%%%%%%%%%%%%%%%%%%%%%%%%%%%%%%%%%%%%%%%
\section{Cosmological hierarchies}  

One of the boundaries of spacetime in the cosmological solution
(\ref{s14}), which will be identified in the following with our brane universe, is
moving with a constant velocity $X = v_1 T$. Again, due to the
cosmological nature of our solution, it is important
in the following to distinguish between the string and the Einstein
frame. 
%%%%%%%%%%%%%%%%%%%%%%%%%%%%%%%%%%%%%%%%%%%%%%%%%%%%%%%%%%%%%%%%%%%%%%%%%%%%%%%%%%%%%%%%%%%%%%%%%%%%%%
\subsection{String frame}

By defining the proper
time on our brane universe
\be
\tau \ = \ {4 \sqrt{1-v_1^2} \over 5 v_1 |q_0| {\kappa}^2} \ ( G_0+ {3 |q_0|
{\kappa}^2 \over 2 } v_1 X_0)^{5 \over 6} \ , \label{h1}
\ee
the induced metric, the overall radius of the five-torus and the
string coupling on our brane have a time dependence governed by
\ba
ds_4^2 &=&  ({{5 v_1 |q_0| {\kappa}^2 \over 4 \sqrt{1-v_1^2}} \ \tau})^{-{2
\over 5}} \delta_{\mu \nu} dx^{\mu} dx^{\nu} - d \tau^2 \ , \nonumber \\
R_c &=& ({{5 v_1 |q_0| {\kappa}^2 \over 4 \sqrt{1-v_1^2}} \ \tau})^{-{1 \over
5}} \ r_c \quad , \quad e^{\Phi} \ = \  \ ({{5 v_1 |q_0| {\kappa}^2 \over 4
\sqrt{1-v_1^2}} \ \tau})^{-1} \ . \label{h2}
\ea
In the string frame therefore the time evolution describes a
contracting universe. Notice that the four-dimensional Yang-Mills coupling in (\ref{s15}),
given by
\be
{1 \over g_{YM}^2} \ = \ e^{- \Phi } \ {R_c^5 \over g^2} \
\sim \ {\rm v}_5  \ , \ \label{h3}
\ee
where ${\rm v}_5$ from now on denotes the dimensionless five-volume
in string units, is time-independent due to the correlation between the time
variation of the string
coupling and that of the internal space. 

On the other hand, the four-dimensional Planck mass $M_P$ and the
effective size $R_X$ of the $X$ coordinate are given by
\ba
M_P^2 &=& {\rm v}_5 M_s^3 \int_0^{v_1 X_0} dX \ e^{-2 \Phi} (G_0+ {3 |q_0| {\kappa}^2 \over 2 }
|X|)^{-{4 \over 3}} = {{\rm v}_5 M_s^3 \over 2 |q_0| {\kappa}^2} \ \biggl[ ( G_0+ {3 |q_0|
{\kappa}^2 \over 2 } v_1 X_0)^{4 \over 3}-G_0^{4 \over 3}  \biggr] \ ,
\nonumber \\
R_X &=& \int_0^{v_1 X_0} dX \ (G_0+ {3 |q_0| {\kappa}^2 \over 2 }
|X|)^{-{1 \over 6}} = {4 \over 5 |q_0| {\kappa}^2} \ \biggl[ ( G_0+ {3 |q_0|
{\kappa}^2 \over 2 } v_1 X_0)^{5 \over 6}-G_0^{5 \over 6}  \biggr]
\ . \label{h4} 
\ea

As the internal five-torus shrinks, there are two possibilities which
must be discussed separately:

i) $r_c >> M_s^{-1}$. In this case, the internal volume starts from
large values and shrinks. We will discuss the hierarchy driven by the 
evolution until the internal radii become of the order the fundamental string
length. In this case, the gauge couplings are tiny and time
independent. This case corresponds to the scenario proposed in
\cite{ap}. 

ii)  $r_c \sim M_s^{-1}$. In this case, T-dualities along the five-torus
coordinates must be performed. The T-dual time dependent solution
corresponds to an expanding internal five-torus perpendicular to the D3
branes. The T-dual string coupling $\exp (\Phi') = 1/{\rm v}_5$ is of
order one and fixes also the gauge couplings 
$(1 / g_{YM}^2) \sim \exp (- \Phi')$. Tree-level Yukawa couplings 
and one-loop generated masses are also time independent, even if the
internal moduli are time dependent. 
 
%%%%%%%%%%%%%%%%%%%%%%%%%%%%%%%%%%%%%%%%%%%%%%%%%%%%%%%%%%%%%%%%%%%%%%%%%%%%%%%%%%%%%%%%%%%%%
\subsection{Einstein frame}

In the Einstein frame, the proper
time on our brane universe is defined by
\be
\tau_E \ = \ {3 \sqrt{1-v_1^2} \over 5 v_1 |q_0| {\kappa}^2} \ ( G_0+ {3 |q_0|
{\kappa}^2 \over 2 } v_1 X_0)^{10 \over 9} \ , \label{h5}
\ee
whereas the induced metric, the overall radius of the five-torus and the
string coupling on our brane have a time-dependence governed by
\ba
ds_4^2 &=&  ({{5 v_1 |q_0| {\kappa}^2 \over 3 \sqrt{1-v_1^2}} \ \tau_E})^{1
\over 5} \delta_{\mu \nu} dx^{\mu} dx^{\nu} - d \tau_E^2 \ , \nonumber \\
R_c &=& ({{5 v_1 |q_0| {\kappa}^2 \over 3 \sqrt{1-v_1^2}} \ \tau_E})^{3 \over
80} r_c \quad , \quad e^{\Phi} \ = \  \ ({{5 v_1 |q_0| {\kappa}^2 \over 3
\sqrt{1-v_1^2}} \ \tau_E})^{-{3 \over 4}} \ . \label{h6}
\ea
The spacetime (\ref{h6}) describes an expanding FRW
universe with an equation of state and  Hubble expansion parameter given by
\be
p \ = \ {17 \over 3} \rho \quad , \quad H \ = \ {\dot a \over a} \ = 
\ {1 \over 10 \tau_E} \ . \label{h06}
\ee
Notice that the four-dimensional Yang-Mills coupling
is again time independent, since it is invariant under Weyl rescalings. 
The effective size $R_X$ of the $X$ coordinate in the Einstein frame is
\be
R_{E,X} = \int_0^{v_1 X_0} dX \ (G_0+ {3 |q_0| {\kappa}^2 \over 2 }
|X|)^{1 \over 9} = {3 \over 5 |q_0| {\kappa}^2} \ \biggl[ ( G_0+ {3 |q_0|
{\kappa}^2 \over 2 } v_1 X_0)^{10 \over 9}-G_0^{10 \over 9}  \biggr]
\ . \label{h8} 
\ee
Let us now assume that the time evolution is valid for a large
cosmological time scale. At that time, some unknown dynamics must take
over which stabilises the dilaton and the two relevant moduli fields
$\sigma , g_{55}$ and produces a late time acceleration of the universe. Our main assumption 
is that this dynamics do not change in a significant way the
hierarchies induced by the previous large cosmological evolution we
are putting forward here. For computing the string coupling and the Planck mass we
use the string frame, while we are careful in order to distinguish
between the string frame radius $R_X$ and the Einstein frame
one, related by the Weyl rescaling (\ref{s15}). Since $\sigma$ and
$\Phi$ are actually related by their $X$-dependence in (\ref{s3}) and (\ref{s13})  
we can in an effective way relate the two length scales by powers of the string  
coupling as $R_X \sim \exp (\Phi /3) R_{E,X}$.
Neglecting factors of order one, the relevant equations
for our purposes are then
\ba
R_{E,X} & \sim & v_1 \ \tau_E \ , \ R_{X}  \sim  v_1 \ \tau \ , \nonumber \\
M_P^2 & \sim & {\rm v}_5 \ (v_1 X_0 M_s)^{4 \over 3} \ M_s^{2} \sim
{\rm v}_5 \ (R_X  M_s)^{8 \over 5} \ M_{s}^2 \ , \nonumber \\
e^{\Phi}  & \sim &   \ (v_1 \tau M_s )^{- {1}} \quad , \quad
R_{c}   \sim  r_c  \ (v_1 \tau M_s)^{-{1 \over 5}} \ . \label{h9} 
\ea 
The velocity $v_1 $ appearing in our solution in our
explicit string example is computed from (\ref{s11}), (\ref{s04}) and
is of order one. In this case, by using $M_P \sim 10^{19}
GeV$, taking as an example $R_{E,X} \simeq 10^{-1} cm$ and by also
using the relation $R_c \sim \exp (\Phi/4) R_{E,c}$, we find from (\ref{h9})  
\ba  
\tau_E &\sim&  3 \times 10^{-12} s \quad , \quad e^{\Phi}
\sim  \ 10^{-15} \ , \nonumber
\\
M_s &\sim& 10^{7} \ GeV \quad , \quad R_{E,c} \sim 10 \ r_c  \ . \label{h10} 
\ea
Notice that the quantum gravity effects appear at the 10d Planck mass 
$M_{\star}= \exp (- \Phi / 4) M_s \sim
10^{11}$ GeV. The time evolution gives rise to a four-dimensional Univers of a mm
size, a new dimension $X$ with an effective mm size as well, five
very small compact dimensions with a very slow time evolution, and
a string coupling becoming tiny, of order $10^{-15}$.   
This conclusion only applies to the case i) of the previous subsection. 
This scenario is similar to the one proposed in \cite{ap}. In the case  
ii) of the previous subsection, even if in the Einstein frame all
internal coordinates are expanding, the string frame analysis tells us
that we must perform T-dualities along the five-torus in order to have a
reliable low-energy effective action description.  
After T-duality, all internal coordinates are perpendicular to the D3
branes, the string coupling and gauge and Yukawa couplings are time
independent and of order one. This case gives a cosmological realisation of
the large extra dimension scenario \cite{add}, which has the additional
interesting feature of keeping time-independence for brane observables
by a time-independent string coupling.
  
There is a second logical possibility we would like to consider, even
if not well motivated by our string examples.  
Namely, at the effective field theory level we can consider the velocity $v_1$ as a continuous
parameter. Then by taking a very small velocity we can accomodate a
much larger time-evolution , until a time prior but close to
the nucleosynthesis. In this case, the model generates three space
dimensions much larger than the others, one moderately large dimension
transverse to our world $R_{E,X} \sim 1 {\rm mm}$ and five small
internal dimensions with a very slow time evolution. The numbers 
corresponding to case i) for this situation are  
\ba  
\tau_E &\sim& 3 \ {\rm mins.} \ \quad , \quad  v_1 \sim 10^{-14} \quad , 
\quad e^{\Phi} \sim  \ 10^{-15} \ , \nonumber
\\
M_s &\sim& 10^{7} \ GeV \quad , \quad  \quad R_c \sim 10 \ r_c \ . \label{h11} 
\ea 
Case ii) realises cosmologically, as before, the large extra dimension
scenario \cite{add}. 
Very small velocities  $v_1 \sim 10^{-14}$ in a string context are
unnatural and hard to get, nevertheless not impossible to realise.  
 
In the case i) and if the solution is valid until nucleosynthesis
there are possible effects coming from unstable but
long lived closed string oscillators, which can decay into (standard
model) fields. Standard Model degrees of freedom in this case must be realized
nonperturbatively, see \cite{ap}. Lifetime of closed string oscillators
is qualitatively of the order of magnitude $\tau \sim \exp (-2 \Phi)
M_s^{-1}$. For a string coupling  smaller than $10 ^{-24}$ or so, they
could in principle be relevant for the dark matter present in the
universe. If the string coupling is however larger, as it is
in our case, we must on the contrary insure that they decay before
nucleosynthesis. This would impose in our case $\exp ( \Phi) \ge 
10 ^{-12}$. 
A more detailed analysis in this case is clearly needed in order to check the
validity of these crude estimates.  
On the other hand, if the dilaton mass is generated by string loop effects,
it will be naturally small and could eventually generate deviations
from the equivalence principle \cite{damour}. 
      
%%%%%%%%%%%%%%%%%%%%%%%%%%%%%%%%%%%%%%%%%%%%%%%%%%%%%%%%%%%
\section{Probe brane in the time dependent geometry}

One of the important questions concerning the 
cosmological solution
found in \cite{dmt} and reviewed 
in Section 2 is its classical
stability. 
One aspect of this is the dynamics of the branes 
which we assumed to be confined on the boundaries.
Namely, we must check that a small perturbation on the moduli
describing the positions of the branes will remain small  with
time.

Let us therefore consider the dynamics of 
a probe brane in the static bulk metric 
\be
ds^2 =  
\biggl[ G_0+ {3{\kappa^2} |q_0| \over 2} X   \biggr]^{1 \over 12} \biggl[  
\delta_{\mu \nu} dx^{\mu} dx^{\nu} -dX_0^2+ dX^2 \biggr]
\ . \label{p1}   
\ee
In the following we define
$\Omega = \biggl[ G_0+ {3{\kappa^2} |q_0| \over 2} X   \biggr]$.
The background dilaton and RR fields are 
\be
\Phi =  \ \Omega^{- {5 \over 6}}
\ ,  \ A_9 = - \Omega^{- {2 \over 3}} \ d^9 x \ , \label{p2}   
\ee
for $X>0$, in a spacetime with boundaries $X=0$ and $X = v_1 X_0$.
A probe brane of tension and RR charge $T_8,q_8$ and of position
$X=X(X_0)$ in this geometry is described by the action
\be
S = - \int_{X(X_0)} d^9 x \ \Omega^{-{2 \over 3}} \ (T_8 \sqrt{1 - {\dot X}^2} \ 
\ - q_8 ) \ .  \label{p3}
\ee
The classical field equation for the position of a probe brane
is then
\be
{d \over dX_0} \biggl( \Omega^{-{2 \over 3}} {T_8 {\dot X} \over \sqrt{1- {\dot
X}^2}} \biggr) \ = \ |q_0| k^2 \ \Omega^{-{5 \over 3}}
\ \biggl( T_8 \sqrt{1- {\dot X}^2} - q_8 \biggr) \ . \label{p4}  
\ee

The hamiltonian of the system described by the lagrangian (\ref{p3})
is
\be
H \ = \ \Omega^{- {2 \over 3}} \ \biggl( {T_8 \over \sqrt{1-{\dot X}^2}}
- q_8 \biggr) \ . \label{p5}
\ee
The equations of motion (\ref{p4}) are valid as long as the brane
is in the bulk.
When the brane arrives at the boundary located at $X=0$ it gets
reflected without changing its energy. When
the brane arrives at the other moving boundary it 
is reflected and boosted by the operation $\Pi_X {\cal K}_{2 \pi \lambda
R}$ so that if its velocity just before the collision
is $u\equiv th(\zeta)$, it becomes just after the collision
\be
u' \ = \ {V-u\over{1-uV}} \ = \ th(2\pi\lambda R-\zeta) \ . \label{p05}
\ee
The energy of the brane $H=E$ is conserved between two
collisions with the moving boundary. The energy changes at the
collision as
\be
{E' \over E} \ = \ { T_8 \ ch(2\pi \lambda R-\zeta)- q_8 \over
T_8 \ ch(\zeta)- q_8} \ .
\ee
Notice that the energy remains invariant for $\zeta=\pi\lambda R$,
i.e. if the brane moves with the same velocity as the boundary.
 
A BPS probe brane at rest ${\dot X}=0$ is a solution of the equations
of motion if the boundaries are static, 
situation that describes the supersymmetric case. 
In our case, it also means that D-branes on top
of the zero-velocity O-planes (corresponding to O-planes at $y=0$ in
the orientifold picture) will remain at rest if their initial velocity
was zero.
 
Solutions of the classical field equations (\ref{p4}) 
have constant
energy $H = E = {\rm const.} \ > \ 0$ between two collisions with
the moving boundary or, equivalently,
\be
{T_8 \over \sqrt{1-{\dot X}^2}} - q_8 \ = \ \Omega^{2 \over 3} \ E \ . \label{p6} 
\ee 
Equation (\ref{p6}) can be analytically solved. For example, in the
probe brane case $T_8 = q_8$, by defining the new
variable
\be
z = \bigl[ 2 + {E \over T_8} \Omega^{2 \over 3} \bigr]^{1 \over 2} \ , \label{p06}
\ee
the solution of (\ref{p6}) is given by the algebraic equation
\be
z^3 - 3 z =({E \over T_8})^{3 \over 2}  {3{\kappa^2} |q_0| \over 2}
(X_0-X_0^{(0)}) \ . \label{p07}  
\ee
The solutions with nonzero velocity satisfy the equation
\be
{T_8 {\ddot X} \over ({1-{\dot X}^2})^{3 \over 2}} \ = \ |q_0| {\kappa}^2 
\Omega ^{-{1 \over 3}} E \ > \ 0  \ . \label{p7}
\ee
The probe branes with nonzero initial 
velocity have a positive acceleration which
will push them on top of the non-BPS system. In
the orientifold picture, this means that branes at the origin 
$y=0$, having zero initial velocity in the $(X_0,X)$ system, remain at the
origin $y=0$ (remain at rest in the $(X_0,X)$ system). Branes in any other
point of the compact space have a net velocity and therefore will reach
in a finite time the $y=\pi R$ nonsupersymmetric system. 
There they will be reflected and boosted
and their fate will depend on the initial velocity they had.
One possibility, if their velocity is negative after the
collision, is that they continue their trajectory until they reach the
other boundary. The other possibility is that due to their
positive acceleration they get reflected once again after some time on the moving brane.
In particular, if initially the velocity was close enough to the
one of the boundary $X = v_1 X_0$, the brane will leave the boundary
for a short period of time and then due to its positive acceleration,
it will collide with it again. After the second collision, we find by using
(\ref{p05}) that the velocity becomes slightly smaller than the one of
the boundary. Then it is accelerated, soon after will collide
again the moving boundary and the process repeat itself. Let us denote
by $v_n$ the velocity of the probe brane immediately after the $n^{th}$
collision, occuring at the time $T_n$ and by $v'_n$ the velocity
immediately after the $(n+1)^{th}$ collision, occuring at the time $T_n
+ \Delta T_n$. Then, if $v_n = (1-\eta_n) th (\pi \lambda R) $ with $\eta_n <<1$, then
between the two collisions the probe brane trajectory can be
approximated with a constant acceleration trajectory provided that
$\Delta T_n << T_n$. The probe brane trajectory in this case is
approximately given by
\be
X (t) \simeq {1 \over 2} a_n (t-T_n)^2 + v_n \ (t-T_n) + th (\pi \lambda R) \ T_n \
, \label{p8}
\ee  
where the acceleration is given by $a_n = {\kappa^2} |q_0| (E/T_8)
\ \Omega (T_n)^{-1/3} (ch (\pi \lambda R))^{-3}$. This intersects again the moving
boundary at $T_n + \Delta T_n$, which gives $a_n  \Delta T_n = 2 \eta_n
th (\pi \lambda R)$. Then it can be shown that $v'_n = (1+\eta_n) th (\pi \lambda R)$ and that
$v_{n+1} = v_n$. This allows us also to estimate the effective time-dependence 
of the energy parameter $E$ at late time $T_n$ as $E_n \sim T_n^{-2/3}
\sim  \Omega^{-2/3}$ which explain, by using (\ref{p6}) why the velocity of the probe brane
can stay very close to the constant velocity of the boundary. By
taking the limit $\eta_n \rightarrow 0$ we see that a probe brane on
the top of the moving boundary will remain there, result that
establish the stability of the initial configuration we assumed in
(\ref{s1}). The analysis and this conclusion is also valid for probe
antibranes. 
An interesting fact is that, once arrived on top of the moving O-planes $X = v_1 T$, the probe brane 
experience a net force (non-zero acceleration) towards the O-plane and
therefore undergoes oscillations around the O-plane trajectory $X = v_1
T$. This is probably interpreted as energy radiated by the probe
brane. A string theory description of these oscillations would be very interesting.

We end this letter with a comment on the possible role played by the
probe brane position to late cosmology (see also \cite{burgess} for
a more detailed analysis of this possibility in brane-world
models. For applications to inflation, see e.g. \cite{kachru}). In the nonrelativistic limit
$\dot X <<1$, the lagrangian (\ref{p3}) becomes
\be
{\cal L}_p = {1 \over 2} \ {\dot \chi}^2 - {c \over \chi} \quad ,
\quad {\rm where} \quad 
\chi = {\sqrt{T_8} \over \kappa^2 |q_0|} \ \Omega^{2 \over 3}  
\ee  
is the canonically normalized brane position and the constant $c =
(T_8-q_8) ({\sqrt{T_8} / \kappa^2 |q_0|}) $ is different from zero for
probe antibranes $T_8=-q_8$. This inverse power-law
potential is typical for scalar fields used in quintessence-type
models of dark energy. In our case, this should happen at very late
time and therefore after the stabilisation of the bulk moduli 
$\Phi, \sigma$ and $R_X$. Our cosmological time evolution plays a
useful role in this respect in ensuring naturally $\chi >> 1$ by time
evolution and therefore providing a very flat $\chi$ scalar potential. 
 
%%%%%%%%%%%%%%%%%%%%%%%%%%%%%%%%%%%%%%%%%%%%%%%%%%%%%%%%%%%%%%%%%%%%%%%%%%%%%%%%%%%%%%%%%
\noindent
{\bf Acknowledgements.} 
We thank Ignatios Antoniadis for useful discussions.
The work of E.D. and J.M. was supported in part by the RTN European Program HPRN-CT-2000-00148. 
%%%%%%%%%%%%%%%%%%%%%%%%%%%%%%%%%%%%%%%%%%%%%%%%%%%%%%%%%%%%%%%%%%%%%%%%%%%%%%%%%%%

\end{document}